# Information Carriers and Identification of Information Objects: An Ontological Approach

**Martin Doerr**[1] and **Yannis Tzitzikas**[1,2]

[1]Institute of Computer Science, FORTH-ICS
[2]Computer Science Department, University of Crete
Heraklion, Crete, GREECE
Email: {martin|tzitzik}@ics.forth.gr

*Abstract:* Even though library and archival practice, as well as Digital Preservation, have a long tradition in identifying information objects, the question of their precise identity under change of carrier or migration is still a riddle to science. The objective of this paper is to provide criteria for the unique identification of some important kinds of information objects, independent from the kind of carrier or specific encoding. Our approach is based on the idea that the substance of some kinds of information objects can completely be described in terms of discrete arrangements of finite numbers of known kinds of symbols, such as those implied by style guides for scientific journal submissions. Our theory is also useful for selecting or describing what has to be preserved. This is a fundamental problem since curators and archivists would like to formally record the decisions of what has to be preserved over time and to decide (or verify) whether a migration (transformation) preserves the intended information content. Furthermore, it is important for reasoning about the authenticity of digital objects, as well as for reducing the cost of digital preservation.

## 1. Introduction

We call '*information identity question*' the problem of deciding whether two *information carriers* (papers, digital files) have identical content, i.e. whether they carry the same *information object*. The identity question seems relatively simple to answer for a material object, such as the Mona Lisa or a person. But even for such things, an a priori agreement on the category of the object to be identified is necessary in order to decide, if "we talk about the same thing" [Wiggins 2001]. However the immaterial nature of an information object makes it far more complex. By "material" we mean anything which has a mass in the sense of modern physics. By "immaterial" we mean that precisely the same (identical) object has the potential to be found on multiple material carriers, e.g. a paper, a computer disk or a tattooed body part. Each "realization" [FRBR] on a particular carrier may undergo different, independent changes at any time. Therefore, an information object cannot be identified with any particular portion [Gerstl 1996] of matter of its carriers.

But what is its substance then? We can say that it is a set of features, but which of all the practically unlimited number of features of a particular physical thing make it up? So, in which sense is it the same or not? For digital objects, the answer seems to be trivial at a first glance: there is a unique binary representation, which can be copied around without loss or alteration. But what if we create from a MS Word document an Abobe PDF version? The binary has radically changed, but in many of the practically relevant cases, such change has not affected the intended information and our copyright on it. So, obviously the law has a concept of identity different from binary forms. And finally, if we print it out, there are no bytes anymore, but still we may keep a copyright on it – on what?

For instance, consider we create a digital object *o1* = `article.doc` in MS Word in order to render some information. Suppose that we print it on paper using a printer *pr* and let *o2* be that printout (physical object). Then we scan *o2* using a scanner *sc* and let *o3*= `article.gif` be the resulting image. Then we apply an OCR (Optical Character Recognition) tool *ocr* on *o3* and let *o4* be the recognized text in ASCII format. We need to be able to identify whether *o4* preserves the intended information object(s) originally encoded in *o1*. If, for instance the intended information did not depend on fonts and page dimensions (plain language), it makes no sense to preserve more than the plain text from *o1*, *o2*, *o4* in order to keep the information.

Scientific journal submission guidelines typically prescribe a limited, known set of highlighting features (such as bold, italic, single and double quotes) that make the possible interpretation of the scientific message in the publication independent from all other font forms and sizes. It is hence in general not the case, as frequently claimed by Digital Preservation experts, that the meaningful elements of a document cannot be determined. We rather take the position that the meaningful features are in most real life cases well known to the author, and in relevant cases, such as scientific publishing, even formally prescribed and known to the publisher. The problem is rather to formalize and preserve this information.

In the above example, we are not concerned with the possible errors of OCR tools, bleeding ink, paper distortions, etc. We are interested in the ideal case, the in-principle ability to compare digital and physical carriers for relevant common features of a certain type by a method which is under sufficient quality conditions deterministic. The very existence of OCR tools that achieve *under ideal conditions* 100% accuracy demonstrates that the method can even be formal. The other striking feature is that it is "optical", exactly as human reading. The basic thesis of this paper is that a notion of identity of information object that conforms to legal practice and the intuition of Digital Preservation must be based on an analysis of the *intended sensory impression* rather than the binary form or material embodiment of an information object.

For that purpose we propose an ontology that provides the concepts and definitions necessary to demonstrate the feasibility of objectifying the sensory impression in order to assess the identity of information content, let us call it "Information Carrying Ontology (ICO)". This has the interesting side-effect that such a theory can cover a much wider range of information objects than, for instance, the popular OIO ontology [Gangemi et al, 2005]. The latter is restricted to formal propositions in terms of an ontology, whereas our theory is based on form and can include even some forms of music. We are not interested in the technicalities of sensory recognition, only in its feasibilities and the kinds of conditions under which they can be deterministic or may become deterministic in the future. We are interested in the theoretical impact of the existence of reliable sensory recognition on the identity of information objects and their relationship to different forms of content representation. Several aspects are beyond the scope of this work. For instance, we do not aim at capturing the *behavior* of digital objects, e.g. how a piece of software behaves, or how a database is supposed to behave (e.g. its query capabilities). This exception also pertains to "active" features in documents such as hyperlinks.

The rest of this paper is organized as follows. Section 2 introduces the basic definitions and introduces the proposed ICO conceptual model. Section 3 provides some indicative examples and applications. Section 4 discusses related work, and finally Section 5 concludes the paper, sketches possible further applications and identifies directions and issues for future research.

A more detailed description of the classes and the associations of the proposed ICO model are given in the Appendix.

## *2. Ontological Perspective and Conceptual Model*

In this section, we describe a model that provides ontological distinctions of the kinds of objects, processes and information assets necessary to discuss how to our opinion information object identity relates to material carriers, digital objects and content recognition.

The proposed model is a structurally object-oriented conceptual model. To avoid reinventing a world ontology, we reuse basically the CIDOC CRM in order to refer to all generic classes out of the immediate focus of this paper, but we make also references to some equivalent classes in DOLCE [Masoloi et al. 2001] and OIO [Gangemi et al. 2005]. The CIDOC CRM ontology (ISO 21127:2006) is a core ontology of 80 classes and 132 relations describing the underlying semantics of over a hundred database schemata and structures from all museum disciplines, archives and libraries. It has already been extended by several more specialized domain ontologies we will refer to later and borrow concepts from. We prefer it here as reference framework because the question of identity of information objects has primary applications in (Digital) libraries, archives, museums and related research in the humanities. In contrast, DOLCE has as empirical base the WordNet dictionary and linguistic intuition, which is more distant of our domain.

The model is presented as a series of UML Class diagrams with specialization/generalization links and associations. The symbols “<” and “>” which sometimes accompany the label of an association are used for specifying the direction of reading the label, while the notations “1”,”*”, “1,*” specify the multiplicity of the association ends (just as in UML Class Diagrams). The symbol “/” at the beginning of an association indicates that the association is *derived*[1]. Since the model is too complex to be shown in one picture and to be described in a top-down manner, we first describe the core model and the basic hypotheses about the role of the sensory impression. Then we go into details about information carrying by material carriers and describe the concept of a unique symbol structure. We continue with the model of information carrying by digital objects in limited analogy with physical information carrying. Then we have all prerequisites to discuss identity and ambiguity of information objects in the narrower sense discussed here.

### 2.1 Core Model and Basic Hypotheses

Figure 1 depicts the core model: In brief, an *Information Carrier carries* one or more *Information Objects* (we disregard the case of empty Information Carriers). In order to analyze what that means, we require that each *Information Carrier* can be subject to one or more *projections, i.e.,* processes whose output is one or more intended *Sensory Impressions*. A *Sensory Impression* is defined as a signal (single or multidimensional) which is **analog** in general. Some may be received by technical sensors, some by human senses and some by both. If the Information Object we are interested in is defined in terms of a **finite, discrete** arrangement of symbols, where each symbol belongs to a finite symbol set and is positioned by adequate arrangement **rules**, it is possible to *extract from* the *Sensory Impression* a *Symbol Structure* which is the substance of the carried *Information Object*. This extraction process can be just reading a paper with my eyes in the sunshine, but also a mechanical process such as

[1] E.g. from the value of an attribute *birthDate* we can derive the value of *age*, and for this reason the latter is called *derived* and it is depicted by /age.

OCR. If the extraction process has reproducible results under sufficient quality conditions, we regard that the carrier *carries* this information.

In this paper, we do not (yet) deal with the sense of "carrying information" and the respective identity question of ***analog*** *information objects*, such as the sensory impression of the Mona Lisa, of the celluloid master copy of a movie, of printouts of digital images with a color or pixel resolution less than the digital original or of playing recordings of bird songs. In the following, we also talk about *Discrete Information Objects* as a synonym of *Symbol Structure,* which is a special case of Information Object, in contrast to *analog information objects*.

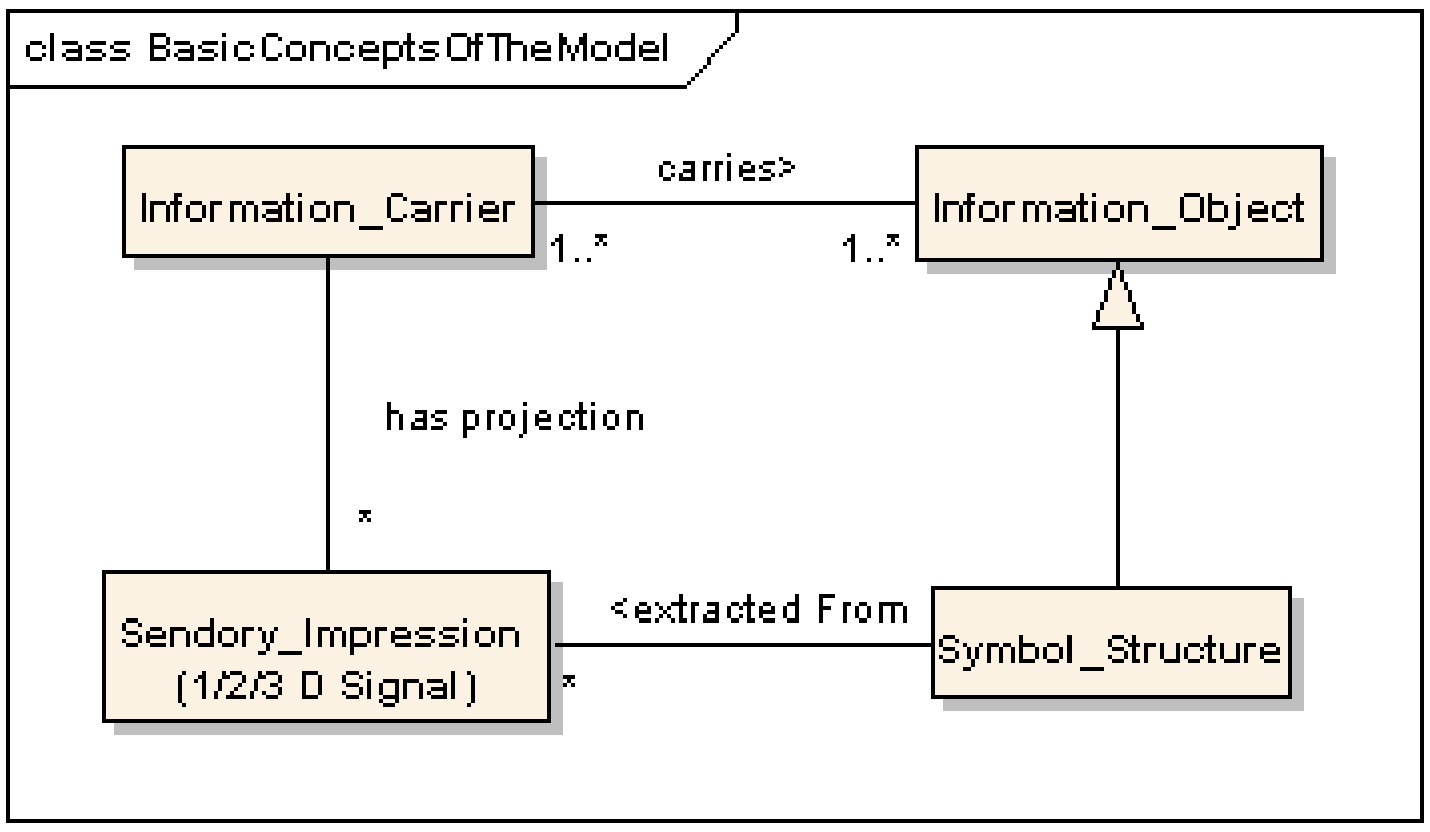

**Figure 1 Some basic concepts of the model**

Regarding the problem of extracting or recognizing an information object on an information carrier, we make the following assumptions:

- The symbol set and the arrangement rules for symbols are a priori known
- We can recognize unambiguously a symbol in the signal (sufficient resolution)
- We can recognize unambiguously the *arrangement* of **all** the symbols in the signal in terms of discrete, meaningful positions.

With the term *arrangement* we refer to notions like: line, begin and end of line, white space, character size (fixed or variable length), direction of reading (e.g. European vs Arabic scripts), etc. We regard that "the (discrete) information object" carried by a carrier is the total of meaningful arranged symbols on that carrier following the given rules, and not any part of it, in agreement with the practice of legal contracts, contents of books etc. We maintain that without a priori knowledge of the symbol set and rules one cannot extract the intended discrete arrangement of symbol occurrences. This knowledge may come from conventions in the context the object was created or be explicitly given by the creator. The idea that this knowledge might be discovered a posteriori from a given carrier is deceiving: Any such case would be based on some cultural assumptions of such rules and the probabilities of their use, which basically is again a kind of priori knowledge. The problem becomes apparent looking at cases of non-resolved archaeological information carriers such as the Phaistos Disc[2]. The process of deciphering texts in unknown scripts is beyond the scope of this paper.

[2] Archaeological object from the Minoan Period found in Phaistos, Crete, Greece, now in the Archaeological Museum of Heraklion.

## 2.2 Detailed Model

In addition to what is shown in Figure 1, we declare a set of formal classes and properties for the concepts mentioned in free text in section 2.1 above and some other auxiliary concepts, in order to concretize our model up to the degree necessary for demonstrating the feasibility of objectifying the sensory impression for defining the identity of information objects. We distinguish in particular the symbolic level of scripts from that of fonts, which is a major source of confusion of identity. This allows us to apply our theory in section 3 to HTML documents in adequate detail. Then we add, in analogy to the chain of properties justifying the property "carries", a chain of properties and auxiliary concepts justifying the transition from digital binary objects to Symbol Structures, which we name "incorporation", in the sense defined in the FRBRoo Ontology [FRBRoo] an extension of the CIDOC CRM approved by IFLA[3] and CIDOC-ICOM.[4] Based on those two major components, we can define objective criteria for the **equivalence** of physical and digital carriers of the **same** information object and formulate the essential properties of information objects. Finally we will detail on the question of ambiguities and disambiguation of information object content and how this may affect the notion of identity of an information object.

The complete presentation of our ICO model is given in a textual form in the Appendix-A. Some of the derived associations of our model are specified also in OCL (Object Constraint Language) in the Appendix-B. The Appendix-A includes some relevant concepts and references to concepts from the CIDOC CRM [LeBoeuf et al. 2012] which overlap with our model or are generalizations of concepts in our model. For each concept the Appendix-A also gives some small examples. More examples that help clarifying the concepts are given in the subsequent sections.

### 2.2.1 Physical Information Carrying

In this section we analyze the sense of material (physical) objects carrying information. A graphical illustration of this "physical model of information carrying" is given in Figure 2: In order to read an *information carrier* in the most general sense we need a *physical projection process* producing from the relevant parts of the carrier a *sensory impression.* Depending on the type of the carrier, such as paper, audio tape, stone surface, vinyl disc, etc., different *physical projection methods* may be used. We distinguish the method intended by the creator of the carrier (*has intended projection*) from the actually used *(used specific technique)*. For instance, the creator of a palimpsest **intended** the pages of the book to be read in day- or candle light, but **use of** infrared projection brings to light the previous, **erased** content. There may be multiple intended methods for one carrier type. For instance, Braille code can be read in daylight and by touching.

From the suitable sensory impression, one may create a *reproduction* of the original carrier by another recording process, which is not analyzed further in this paper (see "F33 Reproduction Event" in [FRBRoo]). For instance, a photocopier would light a document, scan the produced signal and print a paper document with a similar surface color distribution. An analog tape recorder may produce an electrical signal from the tape magnetization and magnetize another

[3] FRBR Review Group of the International Federation of Library Associations

[4] International Committee for Documentation of the International Council of Museums

tape in a similar way by processing the signal. A more indirect way would be to go through the corresponding acoustic signal.

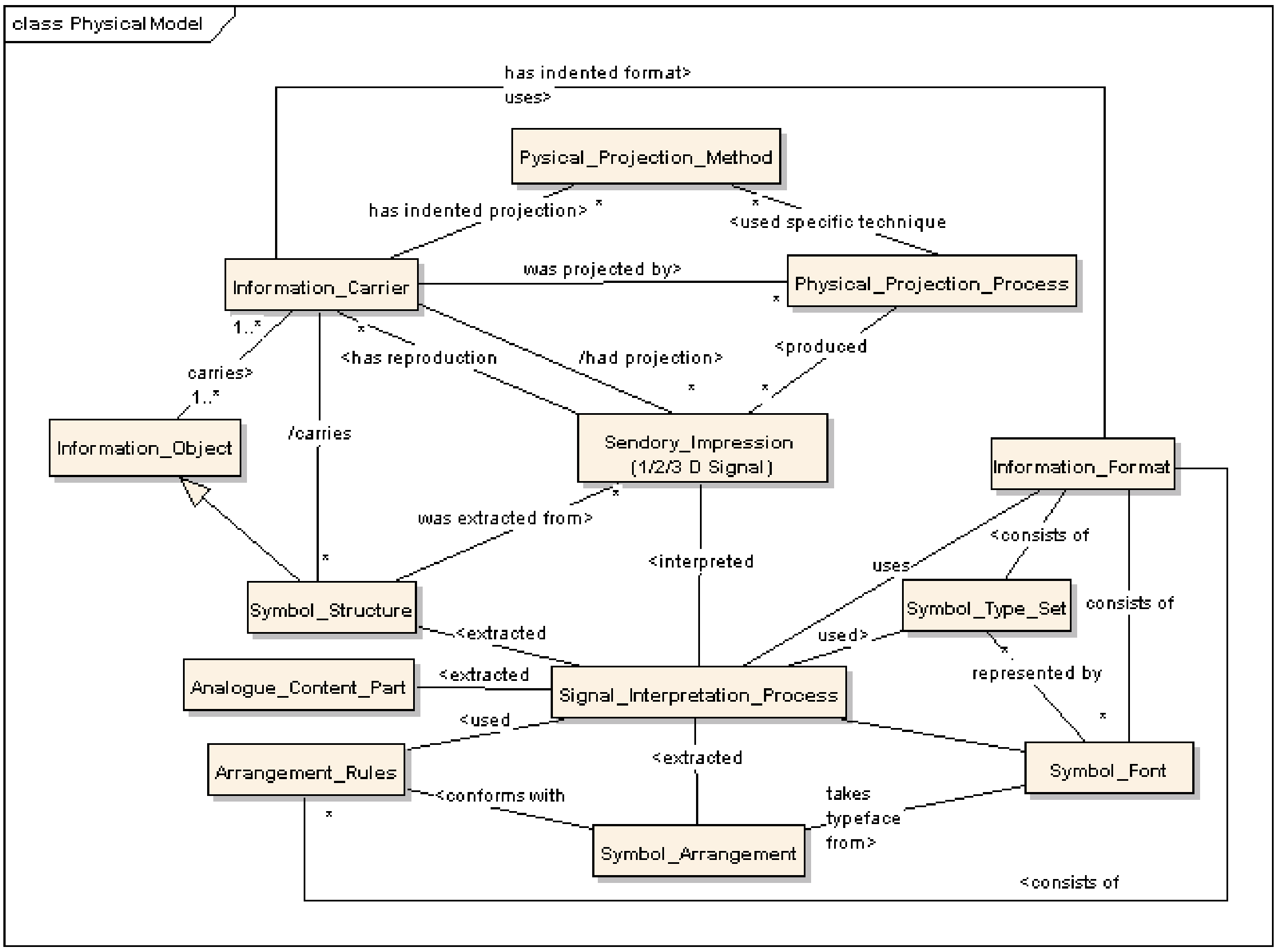

**Figure 2 The physical model of information carrying**

For the projection process, we assume that:

- the physical projection method is sufficient to reliably analyze all intended meaningful information features, such as enough light of suitable color and direction.
- the carrier is **not** deteriorated to a degree that the intended method cannot reliably reveal all intended meaningful information features. If such deteriorations occur, we regard the recoverable information content as **fragment** (see F23 Expression Fragment in [FRBRoo]) of an in general **unknown** information object.
- if reading errors occur, the projection process may be repeated until security about the content is achieved.

However, in order to actually extract the information that we expect to be conveyed by the sensory impression, we need to recognize the information elements in it. This recognition process we call a *signal interpretation process,* which may be human, automated or semiautomatic. Prerequisite of a successful *signal interpretation process* that would recover or "read" the complete intended or relevant information content is the knowledge of the kind of information elements and their syntax within the signal. This in turn is a consequence of the corresponding formatting of the information rendering features on the information carrier when it was "written".

We summarize this knowledge under the term *information format*. It comprises all the formatting information necessary to **identify** the information on a carrier. In case of a rendering of information in discrete symbolic form (a "*discrete information format*"), it must contain the *symbol type sets* used, such as scripts, musical keys etc., the actual symbol prototypes (*symbol fonts*), and the way in which the symbols are arranged (*arrangement rules*) or connected. This applies also to symbolic graphics such as UML or chemical structure graphs. Our concept of information formats also pertains to analog or mixed, discrete and analog information content, but an analysis of components of the latter is out of the scope of this paper. We are not concerned here with format features of a carrier that have no direct correspondence to the intended sensory impression, such as paper thickness. We regard that a "discrete information format" can only describe a *discrete information object*.

As mentioned in section 2.1, we assume that the symbol set and the arrangement rules used to format the information on the carrier are a priori known. In more detail, we distinguish the *information format* actually used by the creator of an Information Carrier ("*uses*") to render the meaningful symbolic or analog information features from the intended semantically relevant format ("has intended format") in the signal. For instance, in order to render plain English language in the format of standard words and syntax on paper, one has to select a character font, but the font is not part of the intended message. To our opinion, this distinction is vital for Digital Preservation, but has been widely overlooked. A source of knowledge about the intended format can be explicit statements of the creator, but also categorical functional criteria, such as texts for legal use in contracts and laws. The effect of this obviously valid distinction is that an Information Carrier may contain more than one information object at the same time, depending on the relevant format definition, as detailed later in section 2.3. Note, that we do not talk here about the parts of the information on the carrier or its parts, but the results of applying different formats to all potential information bearing features on the carrier. Note further, that the information object defined by the **used** information format may not be recognizable due to deterioration, whereas the **intended** one still is.

### 2.2.2 Symbol Structure

We define the unique output of a signal interpretation process of a carrier with a discrete information format as a *symbol structure*, a kind of a graph. It abstracts from all physical features of the carrier and features not related to the intended or sought information content, as qualified by the respective format. For instance, the distance from a paper title to the following section can be abstracted to a "*next in sequence*" link. Any graph structure can be represented by a model of typed nodes and links, for instance, with RDF/S triples. It is the core of our theory, that is **should** be possible for all relevant kinds of discrete symbolic information objects to find a graph structure that uniquely identifies it by **identity in all parts.** The identity of the symbol structure must not depend on any particular encoding. Since we cannot describe it without yet another encoding in a symbolic format, its identity is given by an equivalence relation of different encodings with respect to the graph structure they represent. This equivalence relation is a complete bidirectional mapping. As has been shown by [Zeginis et al. 2011], semantic identity of RDF/S graphs can be defined independent from variations in the serialization. The definition of identity by identity in all parts further implies that an information object in our sense has no versions.

We explicitly include the process of reading digital data carriers, such as tapes, CD, magnetic discs by respective devices in this model. However, the outcome of this process is regarded to be the bit-stream **only**,  the most simple symbol structure possible, and **not** the interpretation of the data as meaningful documents, which will be dealt with in section 2.2.3.

Consequently we assume that different classes of information objects require different models of symbol structures in order to capture their substance adequately. They will be functionally dependent on the respective intended information format. As practical application of our model we propose the creation of specific ontologies of **format features** and the respective symbol structure **models**.

We have no ambition to study such symbol structure models comprehensively in this paper. Inspired from DOM, HTML and XML, scientific paper publishing practice, Roman, Egyptian and Maya inscriptions,  Western music notation practice, modern Chinese and Japanese standard book formats, we **suggest** a generalized form of a *text symbol structure* as shown in Figure 3. It represents a hierarchical structure of components with minimal elements the individual symbols.  There is a notion of sequence of "reading" between those element, and the hierarchical units may overlap, such as pages and paragraphs in books or HTML. XML is actually a labeled ordered tree. We want to demonstrate that at least for these important classes of documents it is possible to define a unique identity of information objects and what the prerequisites are to do so, a notion not sufficiently developed  by Digital Preservation research so far. To which degree artful or poetic editions and expressions can be dealt with by such an approach, future research may show.

In order to develop a practical understanding of what output a *signal interpretation process* in our sense produces, lets us regard a typical reading or OCR process: It would first start with the identification of text lines and text segments consisting of lines, the so-called segmentation process. It would find where the text begins and where it ends. Text segments may have particular semantic roles, such as titles, abstracts, headings. Similarly, in listening to music we would identify the base rhythm. We summarize this knowledge under *arrangement rules*, and the result of the analysis as *symbol arrangement*, in which all relevant relative positions of symbols are turned into the unique semantic relations the positions correspond to. We are not concerned here with artificial examples of minimalistic texts where the notion of a line cannot be found. Following the analysis of the arrangement, the *symbol occurrences* at the respective positions can be recognized and registered. Arrangement interpretation and symbol occurrences together form the *symbol structure*. In addition, the arrangement rules may identify image areas or decorative areas that are integrated in but do not belong to the discrete symbolic content, as *analog content parts*, and register them in the symbol structure. The content of the latter may be saved as it is possible from the signal, or be ignored, but does not form an identification element of the discrete information object. Obviously, chemical formula or UML semantics may need other symbol structure models than those in Figure 3, but yet they all will be based on symbol occurrences and arrangement rules.

Characteristically, in text rendering, we distinguish scripts and fonts. Scripts are a special kind of *symbol type sets*, which can be used to encode and render information, be it natural language texts or any other discrete information content. Symbol type sets must have **finite cardinality**, even small, otherwise messages cannot be deciphered. The classical Chinese script with over 10.000 symbols is probably the largest one humans ever created (decimal number symbols are not unlimited, once they are composed from 10 ciphers). Musical keys can also be seen as symbol type sets. An *information format* would typically restrict the number of *symbol type sets* used in one document. An individual *symbol type* has a meaning and a **prototypical** form. We distinguish the symbol type, such as a Latin "h", commonly called "symbol", from the *symbol occurrence* in a document at a particular position. The forms of symbol types may be varied by a *symbol font*, using bold face, italics, underlines, different sizes and styles. It completely depends on the intended information format, if the distinction between fonts or some fonts is associated with meaning or not, but we regard the symbol type in any case to be associated with meaning. The symbol structure must only register features meaningful with respect to the intended information format (“*has intended format*”).

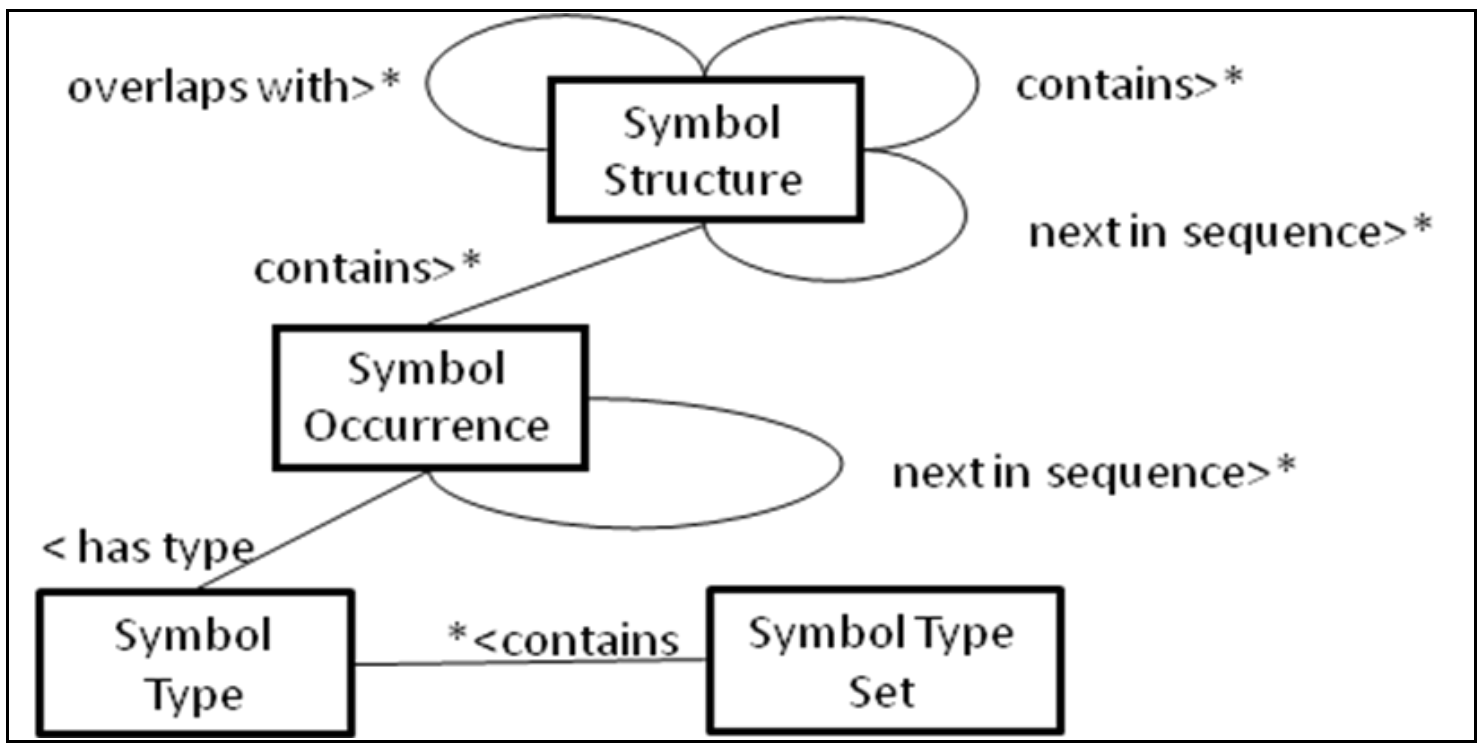

**Figure 3 Symbol Structure**

Future research may show how widely the structure shown in Figure 3 can be applied. In this paper we are interested in demonstrating the feasibility and understanding what the characteristic parameters will be.

### 2.2.3 The Digital Model

Much confusion is caused by the fact that Digital Objects are Information Objects in a sense, but the *intended format* is **not** the binary – the latter is mostly meaningless for human reading. We maintain that similar to physical information carrying, a Digital Object encodes or “carries” information in a sense which is ultimately defined by the sensory impression of the intended rendering methods. A graphical illustration of the Digital model of information carrying is given in Figure 4. In this section we describe our model of analogy and disanalogy of Digital Objects with physical *information carriers* and the information objects carried by them.

A *Digital Object* in our sense is a tuple of a binary content and a format *type*. The binary content is a symbol structure in its own right, which we do not interpret further here. The format type of the Digital Object is the analog of the *intended physical projection method* for an information carrier. The format *type* may encode standardized types such as **mime types,** but in general is specific to the software by or for which it was created. In principle, every new release of some software may correspond to another type of Digital Object it is able to present in the intended way. Only compatibility efforts prevent such a chaos. By executing a viewer or “player”, we call it more generally a *media projection software*, one can launch a *digital projection process*, equivalent to the physical projection process, and create the intended *sensory impression* – in the first place to satisfy a human spectator, reader or listener, as the sensory impression of this Digital Object. As with the physical analog, the *digital projection process* may choose a projection method different from the type of the object, for instance if someone wants to read the binary of a MS Word document for debugging reasons. One may be able to produce a “reproduction” from the sensory impression, either by recording it, or in a **printing** process, in which the intermediate sensory form may or may not be physically used.

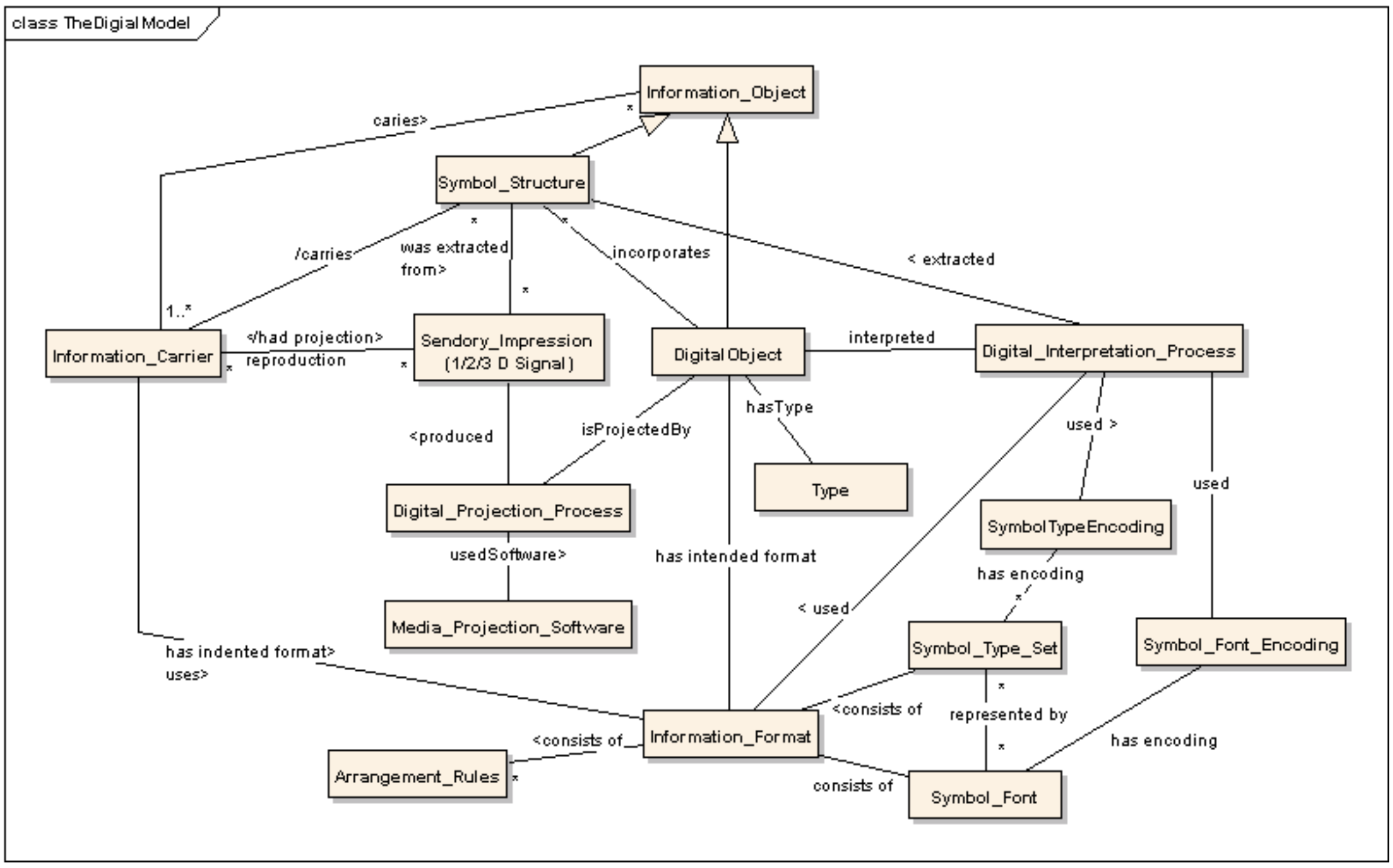

**Figure 4 The Digital Model**

It is the core of our theory that the identity of an information object must be grounded on the intended sensory impression. As with the physical analog, the creator of the Digital Object may have intended the same or another, typically much simpler, *information format* for the interpretation of the content. Encodings, such as PDF, RTF, ASCII etc., as well as font styles, are mostly not part of the intended message, but technically or aesthetically necessary for the representation of the latter. Obviously, with the sensory impression, which is a physical signal, one can execute the **same** *signal interpretation process* as with an impression from a physical carrier and extract the **unique** *symbol structure* which corresponds to the intended information format. We then regard, that the Digital Object *incorporates* a symbol structure, or more generally, an information object.

The relationship "incorporates" was first introduced by [FRBRoo]. It was a major innovation to systematically recognize that an information object may contain other information objects with distinct identity and behavior, without the latter being restricted to a partition or fragment of the containing one. We regard the *incorporation* of the intended information object in the Digital Object as analog to the carrying of an information object by a physical *information carrier*: Since physical reproductions carrying the sensory impression of the information object are possible, the *signal interpretation process* of the digital projection of a Digital Object, of an adequate physical reproduction (printout) of the Digital Object and of a physical reproduction of a printout should result in **exactly** the same *symbol structure* under the same *information format*. Finally, it should also be possible to create another, typically different *Digital Object*, which incorporates the **same** symbol structure. One would call the latter process "migration". This is the ultimate goal of our model: To describe the obvious reality that information objects, such as this text, can exist on paper and in multiple digital forms (PDF, MSWord, OpenOffice etc.), and to explain how and under which conditions this identity can be objectified.

Obviously, for Digital Preservation purposes, one would hardly try to extract the symbol structure incorporated in a Digital Object from the sensory impression itself, once the sensory impression is algorithmically determined. Rather, a *Digital interpretation process* would be used to analyze the binary format from knowledge of how the respective media projection software works. For that purpose, not only fonts and symbol types need to be known, but their actual encoding in the Digital Object binary. Conversion software, such as between MS Word and Adobe PDF format, must be built on an implicit model of the sensory impression by inverse engineering of fonts and arrangement rules.

With this principle of equivalence of Digital Objects and physical information carriers it should be possible, at least for a wide range of types of information objects, to describe Digital Preservation as a preservation of the identity of the information object, independent from the nature of the carrier and the digital format, and not as a fuzzy feature distance as virtually throughout current literature. It should further be possible to finally research and define the **exact** criteria under which such objective definitions of information content are possible, and continuously widen the applicability of this equivalence principle to other content types. In this light, it appears to be particularly relevant to register in the **metadata** not only the used format but the **intended** information format, be it explicitly given or implicit, for instance by the legal functions of documents as they are explicitly known in organizational archival institutions. Attempts to define relevant preservation features for a format category, such as e-mails, rather than for a category of functions, appear rather naive in this light, once a format category can serve any function.

Figure 5 shows the main elements of both (physical and digital) models. Two dotted curves are used for showing the two possible ways to get a symbol structure from a digital object.

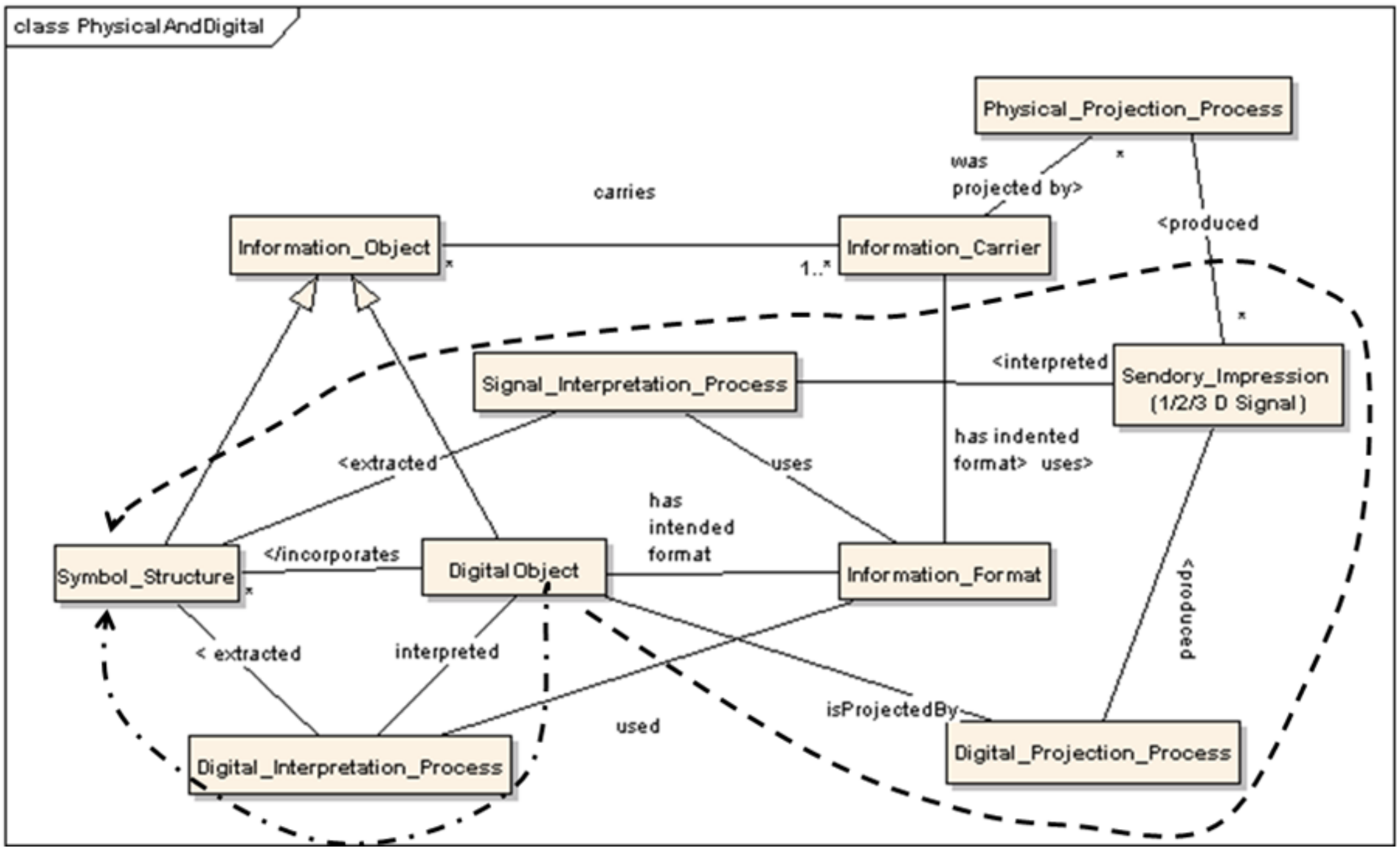

**Figure 5 Physical and Digital Model (part)**

## 2.3 Identity, Incorporation and Ambiguity

As we have seen above, information objects may *incorporate* other relevant information objects. Symbols, such as bits, are used to encode characters, paragraph marks, fonts, etc. Symbols, such as characters, are used to encode words, which are by themselves symbols. Words are used to encode phrases, “propositions”, which may describe “situations” - may be the most abstract form of objective information (see Ontology of Information Objects [Gangemi et al. 2005] in section 4.1). Each such encoding level identifies a new Information Object objectively *incorporated* in another, once it can be specified by a distinct *information format*.

Any Symbol Structure that results from the use of relevant symbol sets and rules in the extraction process which does not conform with the intention of the creator is regarded as identifying a **new** information object carried by the same source. For instance, researchers in critical literature studies may look at the choice of fonts and spread of typos in printed books, marginal later comments or indications of stress in manuscripts. Similarly, the creator him/herself may intend multiple interpretations. Those can be well distinguished, as long as the respective rule sets are known. Guessing what the creator’s intentions were is not our concern here. The higher up we go in the abstraction from the material or binary level, the higher the chance that the **same** information object appears also on another carrier and in another binary. For instance, for a popular work such as Hamlet, it is more likely to find the same set propositions again than the same choice of fonts.

Frequently, physical carriers get corrupted and leave ambiguity with respect to the original information object carried, but also the information object that conforms with the *used* format may be ambiguous by definition. Consider the example shown at Figure 6. Notice that we may not be able to distinguish in Times New Roman the digit "1" at point size 12 from the character "l" at point size 11 (It looks like: 1l). Following our theory above, the use of this font would require a resolution in physical or digital projection and reproduction such that the difference can be resolved unambiguously. So one could say that we need at least 5x4 pixels to discriminate l from 1 (this concerns the concept *Symbol Font* in our model).

However, in most practical cases, we are not concerned with the problem, because, except for things such as passwords, UUIDs and other nonsense strings, there is hardly a message in normal language that would fit to both interpretations, the “l” and the “1”. For instance, suppose that OCR returns "This is a ye11ow cah". We can understand that "ye11ow" corresponds to "yellow" because the latter exists in a dictionary while the former is not. The sequence "cah" also does not correspond to any English word. Two possible choices could be "can" and "cab". We can reject the first choice (the verb "can") since the sentence would not be correct syntactically (however "can" also means "tin"). For many texts, the following sequence of successive interpretation can be used for resolving conflicts: Character, Words, Grammar, Sense, World knowledge. This can be regarded as a general principle of disambiguation: If the information object encodes another level of information, the latter may provide constraints to disambiguate the former. This also holds for error correction systems on Digital carriers.

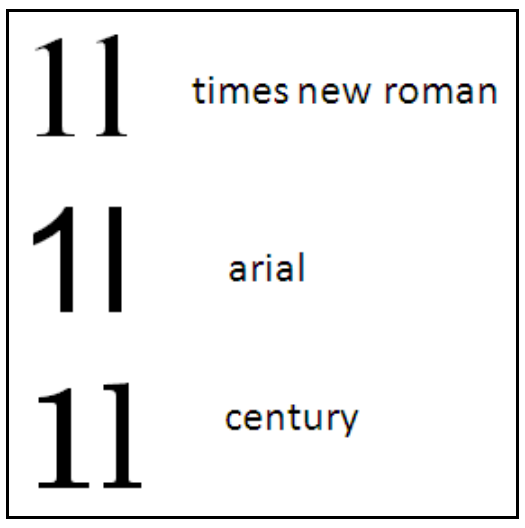

**Figure 6 Examples of symbol extraction ambiguities**

Another case are the letters "U" and "V", which are represented as "V" in the ancient Latin inscriptions. One may regard in such a case that they represent the same Symbol Type ("UV"), rather than as an ambiguous font variant, and in that sense describe the ambiguity as well-defined feature of the information object. In the sequence, occurrences of "UV" can be regarded as spelling variants of the respective normalized Latin words.

Finally, if multiple carriers of the same information object are preserved, one may be able to recover the intended or originally carried information object by comparison of elements ambiguous or lost on one carrier but not on another. Since our definition of identity of an information object is however based on identity in all parts of a symbol structure, an undefined element renders the whole information object undefined, regardless that it might have been intact in the past. Therefore we propose the following rules to deal with identity under ambiguity and information loss:

- Ambiguities about which symbol type occurs or which arrangement feature was used at a particular location and which is a result of successfully applying an intended information format on a non-corrupted carrier is called *systematic ambiguity of recognition*. Any systematic ambiguity that results in a limited set of explicit discrete alternatives is regarded as part of the symbol structure. In that sense, does not affect its identity. The content of the information object is the union of possible interpretations.
- In any other case of ambiguity of recognition of symbols or arrangements we regard the carried information object as **undefined**, in particular after material corruption.
- Due to redundancy with respect to constraints from a more abstract interpretation level, a carried information object may be defined with respect to another, more abstract information format, even though the format used to encode it yields an undefined (partially corrupted) information object.
- Ambiguity of interpretation of symbols must not be confused with the ambiguity of the recognition of symbols and does not affect the identity of the information object in our sense. For instance, the phrase "The rabbit is ready for lunch" has a clear identity as character sequence, even as a correct English phrase, but has ambiguous interpretations.

Based on the distinction of the identity of the carried or incorporated information objects, which depends on the applied information formats, Digital Preservation could develop effective criteria about which interpretation level would be the most **adequate** or the most **effective** to be preserved.

## 3. Examples

This section provides a few examples of how the model can capture some real world cases.

**Example: Three sheets of papers with Latin characters**

Consider three sheets of papers, say *o1, o2, o3,* as shown next.

| **o1:** | **o2:** | **o3:** |
|---|---|---|
| ABC | ***ABC*** | <u>**A**</u><br>***B***<br>**C** |

The following table shows how the conceptual model is instantiated to capture this example.

| *Concept Name* | *Concept instances in this example* |
|---|---|
| **E84 Information Carrier** | Three instances, one for each sheet of paper. |
| **Physical Projection Method** | Lamp with visible light, exposure to daylight |
| **Physical Projection Process** | Exposing the three sheets of papers to daylight. |
| **Sensory Impression** | Three instances, one for each light distribution falling on our retina while looking at these sheets. |
| **Signal Interpretation Process** | Three instances. Each corresponds to the interpretation of one of the three instances of the sensory impressions. |
| **Symbol Type Set** | The Latin characters |
| **Symbol Type** | The Latin characters "A", "B" and "C" . |
| **Symbol Font** | The signal form of the latin characters "A", "B" and "C" in the fonts and styles (italics, underline) which are used in these three sheets. |
| **Symbol Occurrence** | For o1 we have the occurrences "A", "B" and "C", each having the font Currier New.<br>For o2 we have the occurrences "A", "B" and "C" each having the font "Calibri" and is in "Italics".<br>For o3 we have the occurrences "A", "B" and "C" each having font "Times New Roman". Furthermore "A" is underlined and B is in Italics. |
| **Symbol Arrangement** | The arrangement of symbols occurrences. It can be modeled as a graph, like the one shown at Figure 7. |
| **Arrangement Rules** | Instances of rules that specify what a text line is (e.g. sequence of symbols from left to right, etc). |
| **Symbol Structure** | One instance, connected to the instance of the concept Symbol Arrangement. |
| **Information Object** | No need for any instance. The instance of symbol structure is also instance of this class. |

**Table 1 Model instantiation for the three sheets of paper example**

Figure 7 shows a possible modeling of the Symbol Arrangement for the case of *o1*. The diagrammatic notation is that of UML Object Diagrams. Note that there are several methods to model it, actually determined by the way arrangement rules are modeled. In this example we follow a simple modeling approach consisting of an association "next" between symbol occurrences.

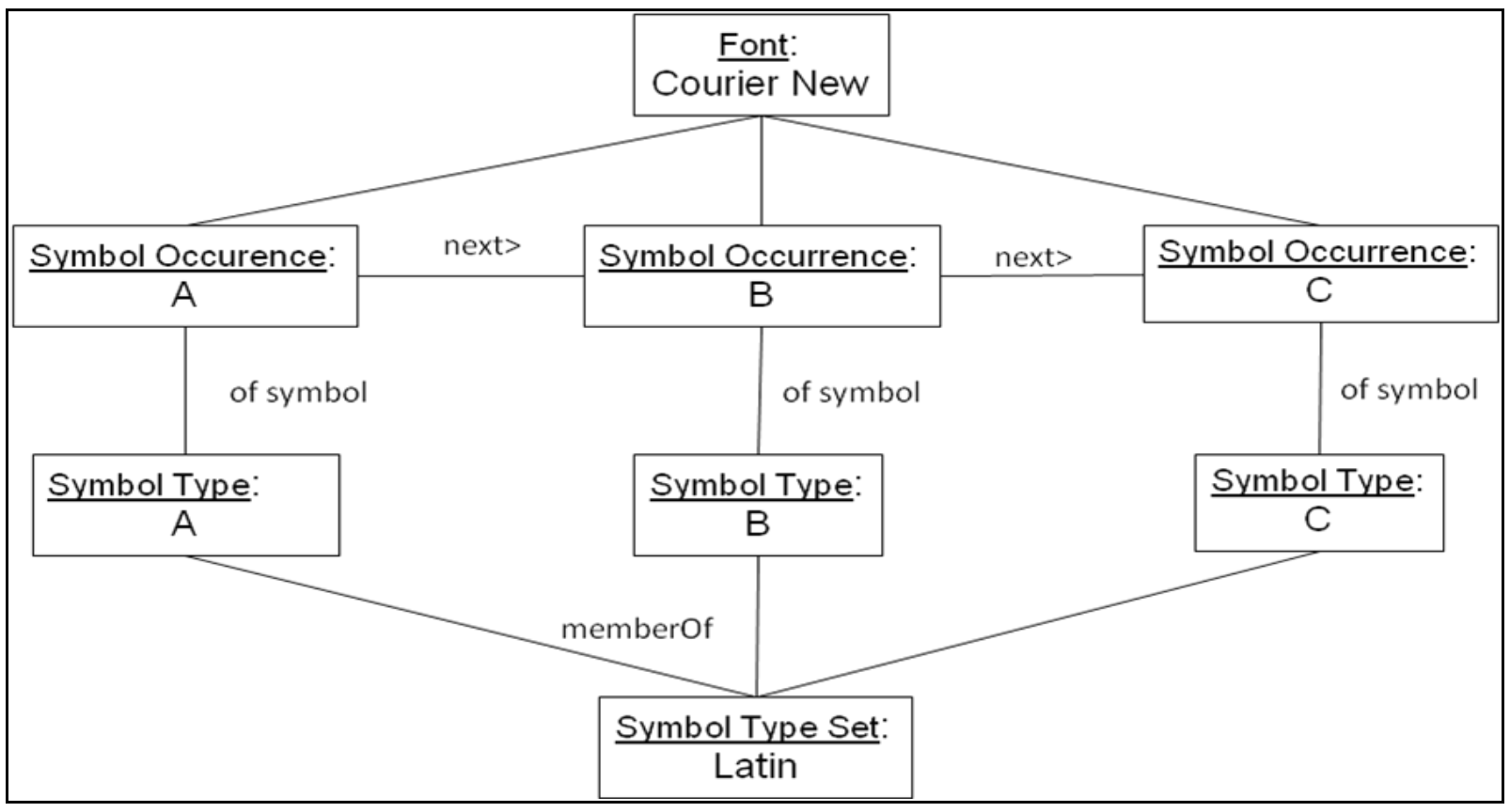

**Figure 7 An example of symbol arrangement**

**Example: An HTML Page**

Here we will describe the case of a digital object, specifically of an HTML file. Let us first recall in brief what HTML is. HTML, stands for HyperText Markup Language, and is the predominant markup language for web pages. HTML is written in the form of HTML elements consisting of tags, enclosed in angle brackets (like <html>), within the web page content. HTML tags normally come in pairs like <h1> and </h1>. The first tag in a pair is the start tag, the second tag is the end tag (they are also called opening tags and closing tags). In between these tags web designers can add text, tables, images, etc. The purpose of a *web browser* is to read HTML documents and compose them into visual or audible web pages. The browser does not display the HTML tags, but uses the tags to interpret the content of the page. HTML allows images and objects to be embedded and can be used to create interactive forms. It provides a means to create structured documents by denoting structural semantics for text such as headings, paragraphs, lists, links, quotes and other items. It can embed scripts in languages such as JavaScript which affect the behavior of HTML webpages. Web browsers can also refer to Cascading Style Sheets (CSS) to define the appearance and layout of text and other material. The W3C, maintainer of both the HTML and the CSS standards, encourages the use of CSS over explicitly presentational HTML markup

The Document Object Model (DOM) is a cross-platform and language-independent convention for representing and interacting with objects in HTML, XHTML and XML documents. Aspects of the DOM (such as its "Elements") may be addressed and manipulated within the syntax of the programming language in use. In brief, each HTML page corresponds to a DOM representation. Figure 8 shows the element types of a DOM tree, while Figure 9 shows a small HTML page and its DOM.

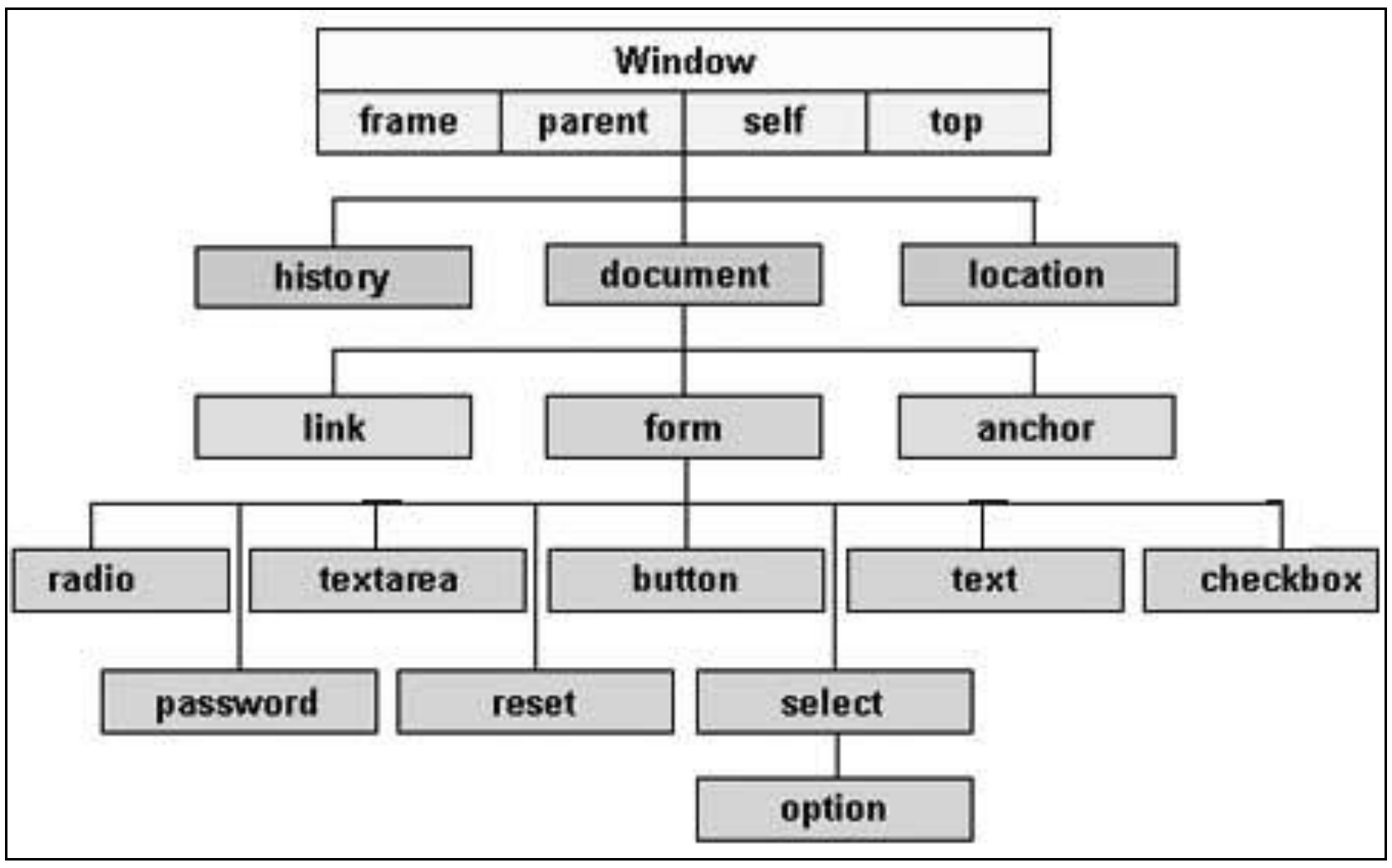

**Figure 8 Element types of a DOM tree**

| **Binary contents** | **HTML contents** |
|---|---|
| 00000000: 3C 68 74 6D 6C 3E 0A 20  20 3C 68 65 61 64 3E 0A<br>00000010: 20 20 20 20 3C 74 69 74  6C 65 3E 0A 20 20 20 20<br>00000020: 20 20 20 20 4D 79 20 54  69 74 6C 65 0A 20 20 20<br>00000030: 3C 2F 74 69 74 6C 65 3E  0A 20 3C 2F 68 65 61 64<br>00000040: 3E 0A 3C 62 6F 64 79 3E  0A 20 20 20 3C 61 20 68<br>00000050: 72 65 66 3D -- 68 74 74  70 3A 2F 2F -- -- 3E 4D<br>00000060: 79 20 4C 69 6E 6B 20 3C  2F 61 3E 0A 20 20 20 3C<br>00000070: 68 31 3E 20 4D 79 20 68  65 61 64 65 72 20 3C 2F<br>00000080: 68 31 3E 0A 3C 2F 62 6F  64 79 3E 0A 3C 2F 68 74<br>00000090: 6D 6C 3E | <html><br><head><br><title><br>My Title<br></title><br></head><br><body><br><a href=”http://…”>My Link </a><br><h1> My header </h1><br></body><br></html> |

| **Sensory Impression (2D Signal)** | **DOM** |
|---|---|
| (see image 1) | (see image 2) |

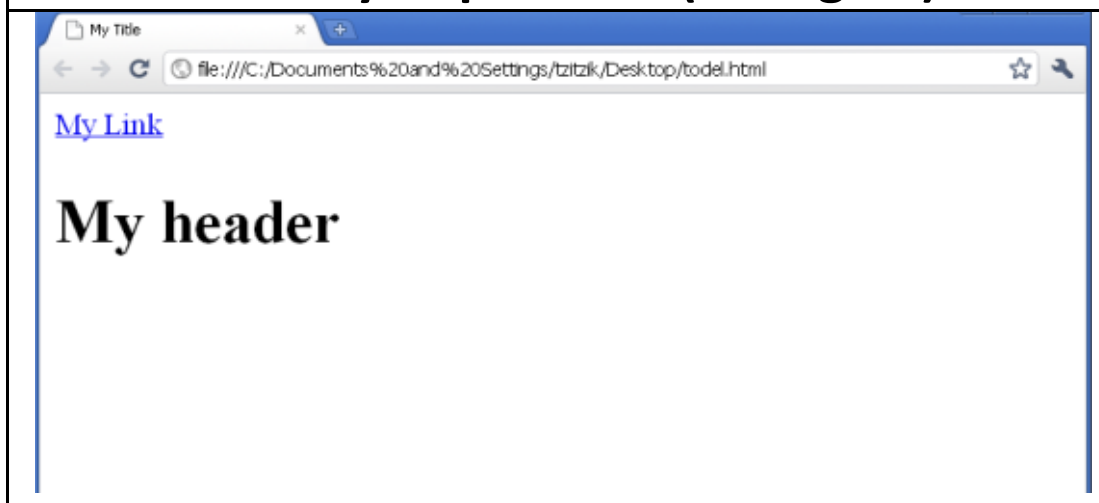

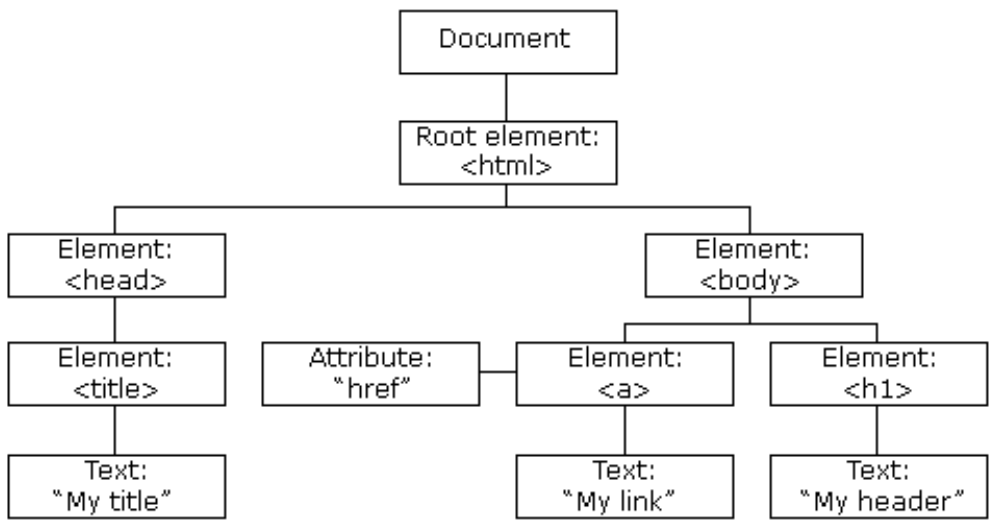

**Figure 9 A toy HTML page and the corresponding DOM tree**

Returning to the problem at hand, consider the case of an HTML page stored as a file on a hard disk in a file named `a.html`. The key point, is that now we are talking about a Digital Object. This means that we do not have to go through *Arrangement Rules* (as in the case of printed sheets of paper). Instead from a Digital Object, the *Digital Interpretation Process* can directly result to a *Symbol Structure* (recall the curves of Figure 5). Note that we would have to go through *Arrangement Rules* if we had not at our disposal the contents of the page in HTML (e.g. if we had a screen dump of the page as rendered by a Web browser). Finally, Table 2 shows how the conceptual model is instantiated to capture this example.

| *Concept Name* | *Concept instances in this example* |
|---|---|
| **E84 Information Carrier** | The hard disk that stores the digital file. |
| **Physical Projection Method** | Hard disk reading? |
| **Physical Projection Process** | The process of hard disk reading (read blocks from the disk) |
| **Digital Projection Process** | The process of showing the contents of the page on a computer screen using a Web browser. |
| **Sensory Impression** | The light distribution falling on our retina when we look at the rendered (by a Web browser) page. |
| **Digital Interpretation Process** | The process of parsing HTML and deriving its DOM representation. |
| **Symbol Structure** | The DOM tree of the page |
| **Information Object** | Instance of the class that is connected the instance of the symbol structure |

**Table 2 Model instantiation for an HTML page**

Since behavioral aspects go beyond the scope of this paper, we will not elaborate on the JavaScripts, as Server-side generated HTML pages, as well as various CSS features.

**Example. Database Contents**
Consider a particular Relational database expressed and managed by a DBMS (e.g. ORACLE). The information carrier in this case is the corresponding blocks in the hard disk. Suppose that for the owner of the database it is sufficient if all contents of the database are preserved in their *logical structuring* and in the form of an ASCII text file (e.g. as a text file that contains one row for each tuple). One can express the *symbol structure* of the database in various formats (from ASCII, to XML or to RDF/S) by exploiting the *conceptualization of the relational model,* e.g. a database is a finite set of relations where a relation is a finite set of tuples, and so on; as defined in the various database textbooks.

We should stress however that we leave out the preservation of the *behavioral aspect* of the database, e.g. the preservation of *query capabilities* of the database. Obviously, a representation in ASCII of the logical structure of the database contents does not directly allow for query answering. Nevertheless, under sufficient functional assumptions, it appears feasible to define a symbol structure for Relational databases that allow for assessing if two databases have identical content, and if identical content is incorporated in different forms of encoding, in particular in back-up media. There is however more work needed, in order to identify the equivalent sensory impression of the content of a database. Clearly, it is out of scope of this paper to speculate how the sensory impression of a database application could be standardized up to a notion of identity. A static notion of database content could be defined in the form of a network of propositions, or lists of records.

**Example. Music and Information Objects**
Let *o1* be the the hand-written sheet music of *Eine kleine Nachtmusik*} (Serenade No. 13 for strings in G major) by Wolfgang Amadeus Mozart and assume that it has a form like the one shown in Figure 10. Let o2 be the digital recording of that piece by the Vienna Philharmonic Orchestra, and let *o3* by a video (e.g. posted to youtube) showing a teenager performing the same melody (e.g. the Violin I part) using an electric guitar. How we can assert that *o2* and *o3* are "authentic" and in what sense? What property of *o1* do *o2* and *o3* preserve?

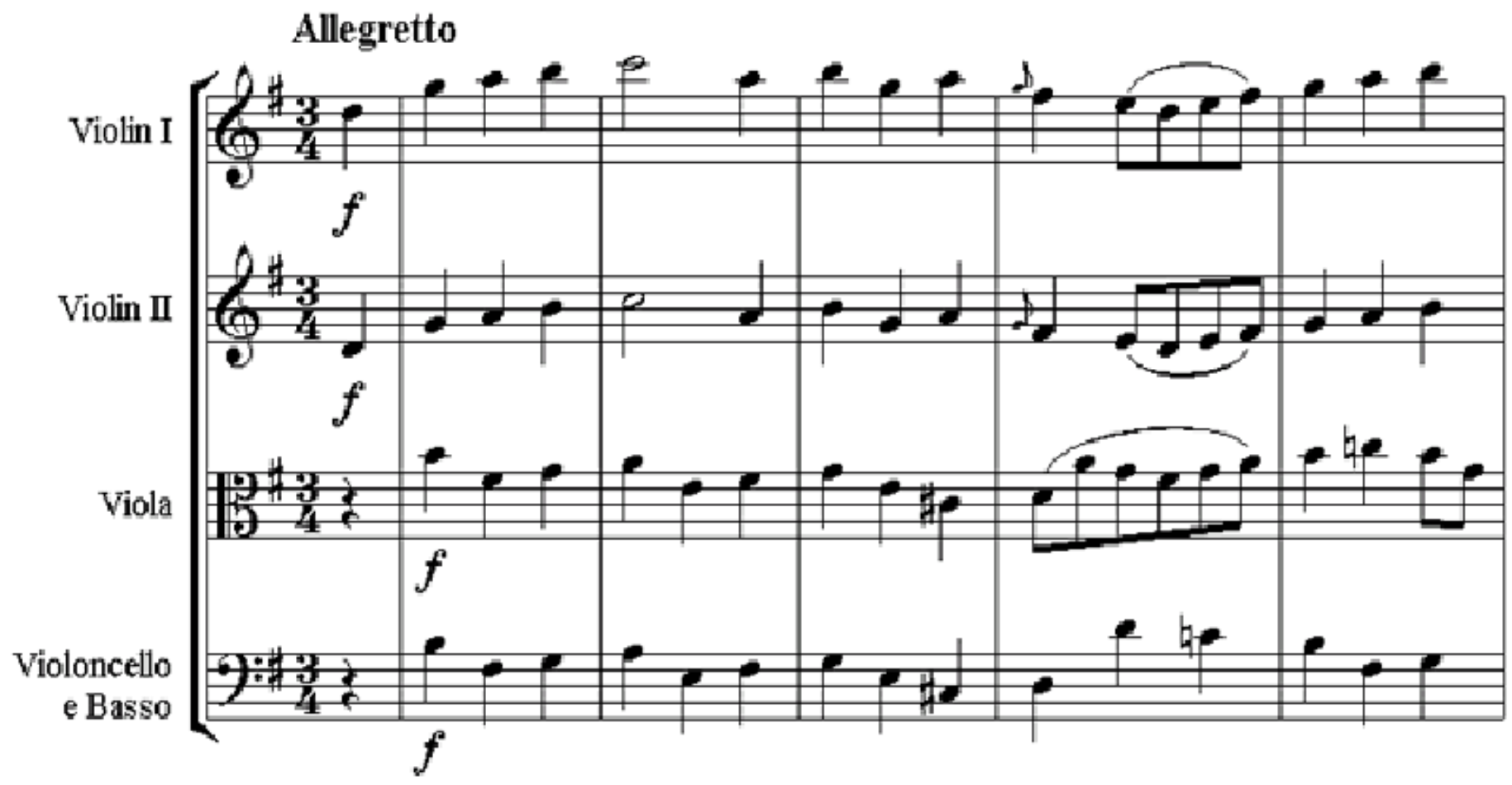

**Figure 10 Musical scores**

Since MIDI allows expressing the basic music data in a standardized format, we could express o1 in MIDI format, and it should be feasible to define a symbol structure in our sense on the basis of the semantic elements in a MIDI representation. There are even tools[5] which can

[5] e.g. Finale (http://www.finalemusic.com)

transcribe scanned pdf score to MIDI, and hence a transition from sensory impression to a symbol structure, which is the key element of our theory. In general, o1 contains basic music information (e.g. key signature, time signature) which can be stored in form of MIDI metadata, and tone information which can be stored in MIDI Channel data. The automatic conversion of the sound channel of o3 to MIDI is feasible. For monophonic music we can extract the series of pitches (e.g by cepstrum or auto-correlation) and hence the duration of each note. By using the pitches we can calculate the key signature, while tempo/beat can be calculated by beat tracking of notes onset. Finally, with tempo and strong/weak beats we can identify the time signature, i.e., the symbol arrangement, while tone can be obtained from frequency spectrum. The automatic conversion of o2 to MIDI is more tough since it is polyphonic music. However we could mention that there are techniques for comparing recorded sound with MIDI for extracting the expressive musical style of recording. To conclude MIDI could be employed for modeling musical information objects and in our example it could be used for checking whether o3 preserves the basic musical features of Violin I part of o1 and ultimately, if o2 and o3 *incorporate* the same information object with respect to a particular definition of relevant features.

Finally, although beyond the scope of this work, we could mention that Gerhard Widmer showed that dynamic/tempo/articulation change are the three atomic features that can describe a pianist's expressive performance [G. Widmer et al. 2004].

**Example. Images (analog information objects)**

This case demonstrates what is possible to do in case we have analogue information objects, i.e. in cases we cannot identify discrete symbols. Suppose we have an image, e.g. the photograph of a drawing. The image can be represented in the feature space as a n-dimensional vector (where n is the number of features). Over that space we can define a metric distance. Each migration of the image to a new format moves the vector of the image. Suppose that each point in time we decide to have stored only one copy of the image. Based on the triangle inequality we can estimate how far the migrated image is from the *original* (initial) image. Analogous analysis can be exploited also for deciding the least resolution that the original photograph of the drawing should have in order to be able to perform a certain number of successive migrations without moving further than a specified distance threshold from the original.

Obviously, on the basis of a distance measure there is **no** notion of identity of the contained information object possible, because the identity relationship must be transitive. Nevertheless, the state-of-the-art of Digital Preservation has not gone beyond distance metrics to describe migration losses (see, for instance, [Becker et al. 2011]), even though we show in this paper that sufficient definition of the relevant features allows for identifying loss-free migration processes for suitable document classes. It is beyond the scope of this paper to investigate, if a fixed maximal distance from a prototype could provide an identity-like relationship of analog information objects for practical purposes.

## *4. Related Work*

We could start by mentioning that Shannon in his theory [Shannon 1948] also assumes that the symbols in a communication session are fixed (and a priori known). This indicates the prominent importance of shared symbols for successful communication (in our case successful and cost-effective preservation).

Below we discuss related ontological approaches and the placement of our work in the context of the general digital preservation strategies and other related works and projects.

## *4.1 Ontological Approaches*

There is no ontology so far that details the meaning of “carrying information” as the one presented in this paper. The most important model of library information, the Functional Requirements for Bibliographic Records FRBR [FRBR] from IFLA, distinguishes the level of meaning (“Work”) from the level of encoding (“Expression”), and the “realization” of an Expression in a physical “Manifestation”. FRBRoo [FRBRoo], an interpretation of FRBR accepted by CIDOC and IFLA, describes the FRBR concepts as a specialization of the CIDOC CRM after some adaptations of the latter.

The CIDOC CRM [LeBoeuf et al. 2012] simply states that physical man-made things can carry information objects (E24 Physical Man-Made Thing. P128 carries: E73 Information Object). It further distinguishes propositional objects, i.e. units of meaning, such as a law in physics, from symbolic objects, i.e., the forms of the encoding signals. An E73 Information Object in the CRM is regarded to have the double nature of a propositional character and symbolic form. The latter distinction was an outcome of the harmonization effort of FRBR with the CIDOC CRM. Whereas the CRM also regards arbitrary physical things, like rocks, to be potential carriers of information, the FRBR view is restricted to physical things produced deliberately for communicating information, as we do in this paper. The CRM prefers the label “carries” rather than “realizes” for the relationship between a carrier and the information object, because linguistically “realizes” implies that the existence AND characteristics of the carrier depend on the preexistence of the information, which is in general not the case.

As all other models, the OIO [Gangemi et al. 2005] model assumes a direct relation “RealizedBy” between the IO and its carrier, but in addition, it recognizes a distinction between the encoded form, the encoding system, a description and the meaning of a description. A *description* in OIO is an instance of a formal ontology that reifies, i.e., talks about, instances of a base ontology in logical terms. OIO develops an elaborate theory about how propositions in a *description* can commit to real-life situations understood in terms of the base ontology. However, OIO does not provide any *identity criteria* for an Information Object that could be operationalized on real documents, because the representation of a *description* is not analyzed. In our model, we detail further what encoding and content means, and how the decoding process takes place. We impose much less requirements on the character of the encoded information than OIO. We only require that the encoded data are reliably identified in terms of known symbols, but not that they conform to a system that can be taken as logical expressions. Otherwise, we would not be able to describe the information content of music or graphics.

The “IRW” ontology [Harpin and Presutti 2009], a proposal to provide a base ontology for the Web architecture, distinguishes *InformationResources* from *NonInformationResources*, which has been taken over in the Europeana Data Model [Doerr et al. 2010]. According to that proposal “*An information resource is a resource whose essential characteristics can be conveyed in a single message. It can be associated with a URI, it can have a representation, for example: a text is an InformationResource*.” and “*All resources that are not information resources*”. A “WebResource”, a subclass of InformationResource, “isRealizedBy” a “WebRepresentation” and isEncodedIn a Mediatype. IRW does not deal at all with physical information carriers others than Webservers, and does not explain the difference in nature of the WebRepresentation versus the WebResource.

## *4.2 Digital Preservation Strategies and Recent Projects*

Most of the works and terms that have come up in the area of digital preservation concern *data*-preservation, not *information*- preservation. For instance, this is true for the terms *bit stream copying refreshing*, *analog backups*, *replication*, etc. One term that is related to information preservation is that of *migration*. It corresponds to a transition to a newer format, operating system or hardware. Resources that are migrated run the risk of losing some type of functionality since newer formats may be incapable of capturing all the functionality of the original format, or the converter itself may be unable to interpret all the nuances of the original format. The latter is often a concern with proprietary data formats. Our analysis can contribute to that problem. Furthermore migration is often performed in a format-based manner. However, and based on our analysis, one could adopt a more fine grained strategy based on what exactly should be preserved.

We should also mention *canonicalization,* which aims at assisting the problem of determination of whether the essential characteristics of a document have remained intact through a conversion from one format to another. Canonicalization relies on the creation of a representation of a type of digital object that conveys all its key aspects in a highly deterministic manner. Once created, this form could be used to algorithmically verify that a converted file has not lost any of its essence. Canonicalization has been postulated as an aid to integrity testing of file migration, but it has not been implemented. According to [Becker 2011], validating the actual content of objects before and after (or during) a preservation action is still one of the key challenges in digital preservation. Our approach can be considered as a concrete method to achieve canonicalization.

The work presented at [Cheney et al. 2001] also aims at understanding the principles of preservation and at modeling the information content representable by physical or digital objects. However the notion of semantics is not analyzed since it is modeled as a function from the objects to the *contents space*, but neither the domain nor the range of that function is analyzed. Essentially that work mainly focuses on the dynamics of preservation and describes it from a quite high level perspective.

We should also mention [Low Jyue Tyan 2011] which presents an extensive review of the literature on "significant properties/characteristics". It concludes that there is a lack of formal objective methodology to identify which characteristics within an information object are significant and should be preserved. Only

***Recent EU Projects*** There are several results from various recently completed or ongoing EU projects. For instance, in PLANETS (FP6 EU project) various metrics for assisting preservation planning were defined (a recent paper is [Becker et al. 2011]). However the indented (for preservation) content has not been modeled, as in many other papers. CASPAR (FP6 EU project) followed the OAIS approach [OAIS] [Giaretta 2011]. Among other things, it attempted to model the notion of *intelligibility* of digital objects according to a *dependency management approach,* and it employed CIDOC CRM (and refinements of CIDOC CRM) for modeling the provenance of digital objects. CASPAR was based mainly on an *encapsulation preservation strategy* that packs together the object plus some extra information. One important kind of such information is what is referred as *RepInfo* which is information about the representation of the data object. The RepInfo can be expressed in languages like EAST [EAST] (or other XML-based languages, e.g. see [Talbott et al, 2006]), while the package can be formed according to various packaging formats like XFDU [XFDU]. Returning to the approach of the current paper, we could identify two "information preservation" approaches: (a) model the information object (as we propose in this paper) and express it in a language (e.g. in RDF/S), (b) leave the original content as expressed in its (carrier) format and add (in its metadata) its "information format", i.e. all required information that is sufficient for getting (a), i.e. for getting the intended information object. Note that EAST is a kind of what we call *information format*, i.e. a language for defining the exact (full) information format of data files. Since RepInfo (or other extra information) may depend on other RepInfo (or other extra and/or external information), the works [Marketakis et al. 2009] and [Tzitzikas et al. 2010] (also in the context of the CASPAR project) elaborate on deciding what to put in a package. In brief this depends on (i) the object, (ii) the assumptions about the designated community, and (iii) the tasks (over the object) that we want to perform in the future. Technically the problem is reduced to *dependency management* and it is implemented using semantic web (RDF/S and rule) technologies. Note that this aspect is orthogonal (or complementary) to the aspect of the current paper (for instance the Information Object, or the Information Format, can be considered and defined to be dependencies of a digital object). Also note that the notion of *task* (and *task-based dependency)* is quite general, e.g. the *projection* that we use in the current paper can be considered as one particular task, or other tasks capturing the *behavioral aspects* of digital files could be defined and managed. In general, that aspect is orthogonal to the aspect of the current paper.

# 5. Concluding Remarks

The basic thesis of this paper is that a notion of identity of information object, that conforms with legal practice and the intuition of Digital Preservation, must be based on an analysis of the intended sensory impression rather than the binary form or material embodiment of an information object. If the information object we are interested in is defined in terms of a finite, discrete arrangement of symbols, where each symbol belongs to a finite symbol set and adequate arrangement rules, it is then possible to extract from the sensory impression a symbol structure which is the substance of the carried information object. For that purpose we proposed an ontology (actually a structurally object-oriented conceptual model that refines CIDOC CRM ontology) that provides the concepts and definitions necessary to demonstrate the feasibility of objectifying the content of a sensory impression to a particular *intended* symbol structure.

The surprising consequence of this model is that an information carrier in general carries more than one information object - depending on the definition of the relevant features applied - and that some of those potential information objects may *incorporate* several other meaningful information objects with less features. This is in contrast to current Digital Preservation Research, which assumes one information content for a material or digital document. Information has a function, it serves communication, and hence has to do with intentions. Therefore it appears quite natural that our model suggests that the definition of information content depends on intention. Objective definition of information content in a use-neutral way appears to us as impossible. We believe that some special applications in literature studies, which may analyze any feature on an information carrier, such as spelling errors or stress of manual writing strokes, are of forensic nature, and should not be confused with information as an object of social function, but are frequently used as argument, that information content cannot be identified. These considerations may explain, why Digital Preservation Research has so much difficulties to identify what has to be preserved.

Under the light of this model we have discussed some indicative real-world examples that motivate the possible range of applicability of this theory and have compared it to related work.

As examples of practical applications, creators of information would like to define which are the contents that when preserved ensure the identity or authenticity of their work. Curators and archivists would like to record formally the decisions of what has to be preserved over time and to decide (or verify) whether a transformation preserves the intended features. In order to use the model presented in this paper in practice, one would be need to define the information formats and respective symbol structures for a series of relevant functional domains, such as scientific publishing in computer science, etc. We expect extensible systems of *feature ontologies* to emerge from such an activity, which could be used by authors and Digital Preservation teams to select for specifying intended information content or content to be preserved. Preservation may support other functions than the authors intended.

There are several further directions and issues that are worth research:

- *Cost-effective Preservation Strategies*. This includes methods and techniques for reducing the amount of information that has to be stored or exchanged on the basis of our analysis (i.e. the intended information object). For example, if the objective is to preserve the piece of text on a sheet of paper, it is a waste to preserve the digital photo of the paper. Note that ``*We create as much information in two days now as we did from the dawn of man through 2003*"[6].
- *Identity and Authenticity.* Our model can be used as a starting point for designing advanced services for identity checking. Specifically it paves the way for equivalence over information contents, and note that different "aspects" of information contents imply different equivalences. *Authenticity* as regards contents, can be defined as a binary relation between two entities that were present at two different events, a current one and a historical one, with respect to a perspective (i.e. information object definition).
- *Behavior Modeling*. Extending the conceptual model for capturing the *behavior* of digital objects (e.g. how a piece of software behaves).

## Acknowledgements

This work was partially supported by APARSEN (Alliance Permanent Access to the Records of Science in Europe Network) , FP7 Network of Excellence, Project number: 269977. Many thanks to A. Mouhtaris for providing us valuable information about the musical example.

---

[6] E. Schmidt (CEO of Google), 2010, http://techcrunch.com/2010/08/04/schmidt-data/

# APPENDIX-A

Formal definition of the ICO Model in an encoding-neutral knowledge representation form. We follow here the format of ISO21127. This definition can be encoded unambiguously in OWL or RDF.

### crm:E73 Information Object

Subclass of: crm:E89 Propositional Object
crm:E90 Symbolic Object

Superclass of: crm:E29 Design or Procedure
crm:E31 Document
crm:E33 Linguistic Object
crm:E36 Visual Item

Scope note: This class comprises identifiable immaterial items, such as a poems, jokes, data sets, images, texts, multimedia objects, procedural prescriptions, computer program code, algorithm or mathematical formulae, that have an objectively recognizable structure and are documented as single units.

An E73 Information Object does not depend on a specific physical carrier, which can include human memory, and it can exist on one or more carriers simultaneously.
Instances of E73 Information Object of a linguistic nature should be declared as instances of the E33 Linguistic Object subclass. Instances of E73 Information Object of a documentary nature should be declared as instances of the E31 Document subclass. Conceptual items such as types and classes are not instances of E73 Information Object, nor are ideas without a reproducible expression.

Examples:

- image BM000038850.JPG from the Clayton Herbarium in London
- E. A. Poe's "The Raven"
- the movie "The Seven Samurai" by Akira Kurosawa
- the Maxwell Equations

### ICI1 Physical Projection Method

Subclass of: crm:E29 Design or Procedure

Scope note: A method of turning the meaningful features on the information layers of an instance of E84 Information Carrier into a signal perceivable by human senses or equivalent sensor devices. Adequate projection methods depend on the nature of the Information Carrier and the way the information has been recorded on it.

Examples:

- The function of a CD player,
- a video tape player function,
- using a lamp with visible light,
- exposure to daylight,
- playing an instrument according to a score.

### crm:E84 Information Carrier

Subclass of: crm:E22 Man-Made Object

Scope note: This class comprises all instances of E22 Man-Made Object that are explicitly designed to act as persistent physical carriers for instances of E73 Information Object.

This allows a relationship to be asserted between an E19 Physical Object and its immaterial information contents. An E84 Information Carrier may or may not contain information, e.g., a diskette. Note that any E18 Physical Thing may carry information, such as an E34 Inscription. However, unless it was specifically designed for this purpose, it is not an Information Carrier. Therefore the property *P128 carries (is carried by)* applies to E18 Physical Thing in general.

Examples:

- the Rosetta Stone
- my paperback copy of Crime & Punishment
- the computer disk at ICS-FORTH that stores the canonical Definition of the CIDOC CRM
- A sheet of paper, a computer hard disc, a hologram, an inscribed stone, a CD-ROM, a videotape, a microfilm.

Properties:

crm:P128 carries (is carried by): crm:E73 Information Object
IP1 has intended projection (is intended projection of): ICI1 Physical Projection Method
IP2 was projected by (projected): ICI2 Physical Projection Process
IP3 had projection (was projection of): ICI3 Sensory Impression
IP31 uses (is used by): ICI19 Information Format
IP32 has intended format (is intended format of): ICI19 Information Format

### ICI2 Physical Projection Process

Subclass of: crm:E7 Activity

Scope note: A process turning the meaningful features on the information layers of an instance of E84 Information Carrier into a signal perceivable by human senses or equivalent sensor devices using a method adequate for the type of Information Carrier in order to reveal its information content.

Examples:

- Playing a score once and completely by a guitar.
- Opening the page of a book in daylight.
- Passing fingers over a Braille code page.

Properties:

IP4 used specific technique (was used by): ICI1 Physical Projection Method
IP5 produced (was produced by): ICI3 Sensory Impression

### ICI3 Sensory Impression

Subclass of: crm:E2 Temporal Entity

Scope note: A physical signal of one or more dimensions. The result of a projection process. A Sensory Impression is a phenomenon bound in space-time characterized by measurable physical quantities. A Sensory Impression may be recorded such that a reproduction of the source carrier can be produced (See also “F33 Reproduction Event” in [FRBRoo].

Examples:

- The light distribution falling on our retina from a page of a book,
- the sound signal of playing a piece of music that reaches a human ear or a microphone,
- the magnetic activation signal reaching a hard disk head.

Properties:

IP27 has reproduction (is reproduction of): crm:E84 Information Carrier

### ICI4 Signal Interpretation Process

Subclass of: crm:E7 Activity

Scope note: The process of receiving and analyzing the information rendered by a signal. Depending on the applied method, technology or attention, the process may recognize symbols, their relative arrangement and analogue characteristics in the interpreted signal.

Examples:

- Reading a book.
- OCR of a paper text.
- Receiving the signal from playing a CD as a bit stream.
- Reading a musical score.

Properties:

IP6 interpreted (was interpreted by): ICI3 Sensory Impression
IP7 extracted (was extracted by): ICI5 Symbol Structure
IP8 extracted (was extracted by): ICI6 Symbol Arrangement
IP9 extracted (was extracted by): ICI7 Analogue Content Part
IP10 used (was used by): ICI8 Symbol Font
IP9 used (was used by): ICI9 Symbol Type Set
IP10 used (was used by): ICI10 Arrangement Rule

### ICI5 Symbol Structure

Subclass of: crm:E73 Information Object

Scope note: The semantic representation of a Symbol Arrangement. We restrict the scope of this class to the cases in which the symbol arrangement can be analyzed in a nested structure of “symbol containers”, which are themselves symbol structures, and may ultimately contain individual symbols, such as chapters, paragraphs, title areas, bold face areas. Both individual symbol occurrences and broader containers may be connected by a set of properties expressing adjacency semantics specific to the arrangement rules intended for the Sensory Impression. Containers may overlap, such as pages and paragraphs.

A symbol structure provides a notion of semantic identity of an (one of several possible) information content of an Information Carrier or Digital Object. It depends only on the ontological commitment of the structural properties and the symbol types, and not on the numerical encoding. For instance, a Latin character “h” is regarded to be the same, independent from its encoding in ASCII, UNICODE, ISO LATIN etc.

For the time being, we do not consider the containment of analogue content parts. In particular, the notion of a Symbol Structure abstracts from any analogue or redundant feature of the symbol arrangement, such as double blanks or extended blanks in justified paragraphs, perspective distortions of digital images of paper documents etc.

Examples:

- The content of a Windows “.txt” file.

Properties:

IP23 contains (is part of): ICI5 Symbol Structure
IP24 next in sequence (previous in sequence): ICI5 Symbol Structure
IP25 overlaps with: ICI5 Symbol Structure

### ICI6 Symbol Arrangement

Subclass of: crm:E73 Information Object

Scope note: This class comprises instances of E73 Information Object which consist of a particular, discrete arrangement of symbol representation occurrences in the arranged space with well-defined position semantics.

Examples:

- The relative positions of all characters on the front page of a news paper.

Properties:

IP11 conforms with (is conformant rule of): ICI10 Arrangement Rule
IP12 takes type face from: ICI8 Symbol Font

### ICI7 Analogue Content Part

Subclass of: crm:E73 Information Object

Scope note: This class comprises segments of instances of E73 Information Object which have an intended structure that does not follow discrete arrangement rules of finite symbol sets.

Examples:

- The color distribution of the Mona Lisa.
- The content of a celluloid movie.
- A press photo published in a newspaper

### ICI8 Symbol Font

Subclass of: crm: E89 Propositional Object

Scope note: A set of Symbol Representations of all Symbol Types of a Symbol Type Set. Different Symbol Types of all used Symbol Sets in an Information Object may not share the same Symbol Representation. A font may have digital identifiers. Normally, each symbol of a Symbol Type set has exactly one, distinguished Symbol Representation in a Symbol Font . We generalize here the notion of “font” from traditional paper printing to any symbol representation in any media signal form.

Depending on the interpretation rules for the intended meaning of an Information Carrier, symbols represented in different fonts may be regarded as different symbols or not. Frequently, maximal contiguous symbol sequences written in the same font represent an area in the symbol arrangement with specific semantics, such as the use of bold face for stressing a word.

Examples:

- Times New Roman Bold Italic (True Type)
- The musical key “do minore” on an Irish harp

Properties:

IP13 represents (is represented by): ICI9 Symbol Type Set
IP14 has encoding (is an encoding of): ICI11 Symbol Font Encoding

### ICI9 Symbol Type Set

Subclass of: crm: E89 Propositional Object

Scope note: A finite set of symbol types used to encode or interpret information, such as a script. All symbol types in a symbol type set must be mutually disjoint. A Symbol Type Set can be represented by multiple Symbol Fonts and multiple symbol type encodings. In a sense, the Symbol Type Set represents the set of equivalence classes of the various symbol encodings in use.

Examples:

- The Latin characters, musical tones,
- The musical key "do minore"
- digital bits 0,1

Properties:

IP15 has encoding (is an encoding of): ICI12 Symbol Type Encoding
IP26 contains (is part of): ICI18 Symbol Type

### ICI10 Arrangement Rule

Subclass of: crm:E29 Design or Procedure

Scope note: This class comprises rules that specify (constrain) possible arrangements (relative positioning) of symbols in a signal to discrete positions that determine semantic characteristics such as symbol order, new-line, subdivisions, page numbers or title positions.

Examples:

- The notion of a text line, character spacing and sets of continuing text lines.
- An kinds of areas that can be covered by a paragraph.
- "Boustrophedon".
- The rythm of a score.
- the duration and frequency levels of a note.
- The set of layout specifications possible in HTML.

### ICI11 Symbol Font Encoding

Subclass of: crm: E73 Information Object

Scope note: A mapping table assigning to each symbol representation a unique identification and an identifier of the respective symbol type taking from one or more symbol type encodings ("script tag").

Examples:

- AGaramondPro-Regular.otf

### ICI12 Symbol Type Encoding

Subclass of: crm: E73 Information Object

Scope note: A finite set of symbol types used to encode or interpret information. All symbol types in a symbol type set must be mutually disjoint. A Symbol Type Set can be represented by multiple Symbol Fonts.

Examples:

- ASCII code Latin character encoding
- ISO Latin-1

### ICI13 Digital Object

Subclass of: crm: E73 Information Object

Scope note: This class comprises data objects characterized by a unique binary content and a type information (such as a “mimetype”). Note that the identity condition of an ICI13 Digital Object relies on the unaltered bit content and type assignment.

Examples:

- Tzitzikas_Doerr_V2.docx, as modified 7/8/2011
- Beatles - Come Together.mp3, as modified 12/9/2008

Properties:

IP16 has type: crm:E55 Type
IP26 incorporates: ICI5 Symbol Structure

### ICI14 Digital Projection Process

Subclass of: crm:E7 Activity

Scope note: Running of media projection S/W or firmware turning the bits of an instance of ICI13 Digital Object into a signal perceivable by human senses or equivalent sensor devices according to the intended meaning of the Digital Object. Adequate media projection S/W depends on the type of the Digital Object.

Examples:

- Showing the content of Tzitzikas_Doerr_V2.docx on my computer screen with MS Word.
- Playing Beatles - Come Together.mp3, as modified 12/9/2008 with Windows Media Player.

Properties:

IP17 used software (was used by): ICI15 Media Projection Software
IP18 produced (was produced by): ICI3 Sensory Impression

### ICI15 Media Projection Software

Subclass of: crm:E7 Activity

Scope note: Media projection S/W or firmware turning the bits of an instance of ICI13 Digital Object into a signal perceivable by human senses or equivalent sensor devices according to the type of the ICI13 Digital Object. In case of visual media, we would speak about “viewers”, in case of audio or video of “media players”.

Examples:

- MS Word in view mode.
- Windows Media Player.

### ICI16 Digital Interpretation Process

Subclass of: crm:E7 Activity

Scope note: The process of analyzing an ICI13 Digital Object in order to extract an incorporated Symbol Structure according to a set of rules. In particular, it may be able to reproduce the Symbol Structure a Signal Interpretation Process would extract from a Digital Projection of the same Digital File, or reproduce from a Digital Image of an Information Carrier the Symbol Structure a Signal Interpretation Process would extract from the same Information Carrier.

Examples:

- “Cut and paste” of a text string from a .pdf file to a MS Word document with the function “keep text only”.
- OCR from a Digital Image of a text page.

Properties:

IP19 interpreted (was interpreted by): ICI13 Digital Object
IP20 extracted (was extracted by): ICI5 Symbol Structure
IP21 used (was used by): ICI11 Symbol Font Encoding
IP22 used (was used by): ICI12 Symbol Type Encoding

### ICI17 Symbol Occurrence

Subclass of: crm:ICI5 Symbol Structure

Scope note: This class comprises occurrences of single Symbol Types in a Symbol Structure. A symbol occurrence is characterized by the type of the occurring symbol and the semantically relevant relative position within the symbol structure, typically to the adjacent symbols or larger unit boundaries. Symbol Occurrences are the minimal elements of Symbol Structures.

Examples:

- The “a” in “class” in the scope note above.

Properties:

IP25 has type (is type of): ICI18 Symbol Type

### ICI18 Symbol Type

Subclass of: crm:E90 Symbolic Object

Scope note: An individual symbol type of a Symbol Type Set used to encode or interpret information. A Symbol Type has one or more names, possible digital representation codes (identifiers) and one or more possible Symbol Representations.

Examples:

- The Latin character “h” .
- The non-smoking sign.
- The musical note “do minore”.

### ICI19 Information Format

Subclass of: crm:E29 Design or Procedure

Scope note: A set of rules and symbol types that govern the intelligibility and symbolic interpretation of the information content in a Sensory Impression intended by an information carrier. In case of a format of discrete, arranged symbols, the Information Format must specify Symbol Type Sets, Symbol Fonts and Arrangement Rules. In case of analogue information it may specify things like color ranges and relevant carrier parts. The format actually used by an Information Carrier is in general much more detailed than the semantically intended one. For instance, in order to render plain English language in the format of standard words on paper, one has to select a character font.

Examples:

- Imperial Roman inscription format
- Springer’s research article format

Properties:

IP28 consists of (forms part of): ICI8 Symbol Font
IP29 consists of (forms part of): ICI9 Symbol Type Set

IP30 consists of (forms part of): ICI10 Arrangement Rule

## APPENDIX-B

The specification in OCL (Object Constraint Language) of the derived associations of our model:

```
Context Information_Carrier::had_projection:Set
derive: was_projected_by.produced

Context Information_Carrier::carries:Set
derive: was_projected_by.produced.was_extracted_from
```

**Figure 11 Specification of the derived associations**